\documentclass[
reprint, 
superscriptaddress, 
amsmath,
amssymb, 
aps,
floatfix,
longbibliography
]{revtex4-1}
\usepackage{graphicx}
\usepackage{xcolor}
\usepackage[colorlinks=true,citecolor=blue,linkcolor=magenta]{hyperref}
\usepackage{bm}

\begin{document}

\title{Nonadiabatic transitions during a passage near a critical point}

\author{Nikolai~A.~Sinitsyn}
\affiliation{Theoretical Division, Los Alamos National Laboratory, Los Alamos, New Mexico 87545, USA}
\author{Vijay Ganesh Sadhasivam}
\affiliation{Theoretical Division, Los Alamos National Laboratory, Los Alamos, New Mexico 87545, USA}
\affiliation{Yusuf Hamied Department of Chemistry, University of Cambridge, 
Cambridge, CB2 1EW, 
UK}

\author{Fumika Suzuki}
\affiliation{Theoretical Division, Los Alamos National Laboratory, Los Alamos, New Mexico 87545, USA}
\affiliation{Center for Nonlinear Studies, Los Alamos National Laboratory, Los Alamos, New Mexico 87545, USA}
\begin{abstract}
   The passage through a  critical point of a many-body quantum system leads to abundant nonadiabatic excitations.  Here,  we explore a regime, in which the critical point is not crossed although the system is passing  slowly very close to it. We show that the leading exponent for the excitation probability then can be obtained by standard arguments of the Dykhne formula but the exponential prefactor is no longer simple, and behaves as a power law on the characteristic transition rate. We derive this prefactor for the nonlinear Landau-Zener (nLZ) model by adjusting the Dykhne's approach. Then, we introduce an exactly solvable model of the transition near a  critical point in the Stark ladder. We derive the number of the excitations for it without approximations, and find qualitatively similar results for the excitation scaling. 
\end{abstract}

\maketitle

\section{Introduction}
The nonadiabatic excitations that emerge during a slow passage through a quantum critical point limit the performance of  quantum annealing computers \cite{yan-ncom}. The theory of such critical excitations has also a broad range of applications, from cosmology to quantum metrology and control \cite{kibble,oleh,del-kampo}. 

For a 2nd order quantum phase transition at zero  temperature, the density of the excitations, $n_{ex}$, after the slow crossing of the critical point is known to follow a power law \cite{del-kampo,altland2009,KZ1,KZ2,jacek}
$
n_{ex} \propto \beta^{\nu},
$
where $\beta$ is the characteristic rate of the transition through the critical point. For the 1st order phase transitions, in the absence of dissipation,  the critical point crossing  leads to a much stronger excited state with $n_{ex}=O(N)$, where $N$ is the number of particles in the system \cite{niu-nlz}. 

This behavior is in conflict with the adiabatic theorem that predicts an exponential suppression of the nonadiabatic excitations in the adiabatic limit. The controversy is explained by the fact that near the quantum critical point the energy gap is vanishing with $N$, whereas the adiabatic theorem assumes that the energy gap is finite.    

Here, we consider a different 
 regime, in which the time-dependent control makes a system pass in the vicinity of the critical point without crossing it. As the energy gap then does not close, it is expected that the nonadiabatic excitations are suppressed exponentially for slow transition rates: $n_{ex}\propto e^{-f/\beta}$, where $f$ is a constant and $\beta$ now is the characteristic rate of the transition near the critical point. Nevertheless, the vicinity of the critical point should enhance the excitations, especially in the regime of a moderately slow passage. Hence, we want to know how the excitation probability depends on the distance of the passage near the critical point.     

This question has been studied before only in the context of the nLZ model. Namely, in \cite{niu-nlz} the exponential suppression of the excitations was derived but without good  agreement with numerical simulations. Among possible reasons for this disagreement, the authors of  \cite{niu-nlz} mentioned a nontrivial exponential prefactor that was likely contributing to $n_{ex}$ in the nearly-critical regime.

In this article, we provide an evidence to that, indeed, the number of the nonadaibatic excitations after the passage near a critical point should depend on the characteristic transition rate $\beta$ as
\begin{equation}
n_{ex}\propto \beta^{\nu} e^{-f/\beta},
\label{scal1}
\end{equation}
which means that the exponential prefactor is nontrivial, and depends as a power law on the transition rate. The constant $f$ depends on the minimal gap that was reached during the time evolution. We will also show that if there is a constant parameter $\kappa$,  such that the critical point would be touched during the time evolution at $\kappa_c$,
then for small $|\kappa-\kappa_c|$ the exponential suppression in (\ref{scal1}) is 
characterized by another type of the power law: $f\propto |\kappa-\kappa_c|^{\mu}$.  Hence, there is usually a significant  range of parameters, such that  the physically interesting values of the excitation probability, $P_{ex}>10^{-3}$ per particle, can be described essentially by the power law prefactor in (\ref{scal1}) rather than the exponent in (\ref{scal1}).  In experiments, as well as numerical simulations, this behavior may be interpreted incorrectly as the result of a transition through the 2nd order critical point.

Finally, our theory extends to the regime when the critical point is merely touched, i.e., when  $\kappa=\kappa_c$. Then, 
we predict a  power law suppression of the excitations: $n_{ex}\propto \beta^{\nu_c}$, where $\nu_c$ is  different from $\nu$ in (\ref{scal1}).

\section{The nonlinear Landau-Zener model}
The nLZ model describes $N$ spins-1/2 with all-to-all Ising-like interactions and coupled to a linearly time-dependent magnetic field:
\begin{equation}
 H(t)=\sum_{i=1}^N (\beta t \sigma_i^z +g\sigma_i^x) - \frac{k}{N} \sum_{i\ne j} \sigma_i^z \sigma_j^z,
    \label{nLZ-def}
\end{equation}
where $\sigma_i^{\alpha}$ are the Pauli operators of the $i$-th spin;  $\beta$ is the sweep rate of the field, and $g$, $k$ are the coupling constants. We assume that $\beta>0$, and consider the limit $N\rightarrow \infty$ that is taken before the quasi-adiabatic assumption, $\beta/g^2 \ll 1$.
We will not discuss specific physical applications of the nLZ model, but only mention here that it emerges in many contexts, including chemical reactions between Bose-Einstein condensates and magnetic hysteresis in molecular nanomagnets in time-dependent fields \cite{niu-nlz}. 

Without the last interaction term in (\ref{nLZ-def}), all spins would experience the standard Landau-Zener evolution with a fixed field $g$ along the $x$-axis and a linearly time-dependent field, $\beta t$, along the $z$-axis. The Landau-Zener model is  solvable, so for $k=0$, the probability per spin to produce a nonadiabatic excitation is known exactly \cite{niu-nlz}:
\begin{equation}
P_{ex}^{LZ} =e^{-\pi g^2/\beta}.
\label{plz}
\end{equation}

\subsection{The mean field approximation}

For finite $k$, the exact analytical analysis becomes impossible, so we consider the limit of a very slow transition: $g^2/\beta \gg 1$.  
Near the critical point and for $N \gg 1$, the all-to-all interactions justify the applicability of the mean field approximation, so that the effective Hamiltonian for the $i$-th spin is given by 
\begin{equation}
  H_i(t) = [\beta t -kS_z(t)]\sigma_i^z +g \sigma_i^x,
    \label{i-nlz}
\end{equation}
where $S_z(t)$ is the mean polarization field of all spins: 
\begin{equation}
    S_z =\frac{1}{N}\sum_{i=1}^{N} \langle \sigma_i^z \rangle.
    \label{sz-def}
\end{equation}

The evolution starts as $t\rightarrow -\infty$ in the ground state with  $\langle \sigma_i^z \rangle =1$ for all $i$. For the adiabatic evolution, as $t\rightarrow +\infty$, all spins must flip to the states with $\langle \sigma_i^z \rangle =-1$. For intermediate time, the adiabatic evolution makes the spin always point against the direction of the net magnetic field that has components  $(B_x,B_y,B_z)=(g,0,\beta t-kS_z(t))$. This gives us the value of $\langle \sigma_i^z \rangle$ in the strict adiabatic limit:
$$
\langle \sigma_i^z \rangle = -\frac{B_z}{\sqrt{B_x^2+B_z^2}} = -\frac{\beta t-kS_z}{\sqrt{(\beta t -kS_z)^2+g^2}}, \,\,\, \forall i. 
$$
Substituting this into (\ref{sz-def}) we find the equation that defines $S_z(t)$ in the quasi-adiabatic case:
\begin{equation}
S_z(t)= -\frac{\beta t-kS_z(t)}{\sqrt{(\beta t -kS_z(t))^2+g^2}}.
    \label{sz-self}
\end{equation}
We consider the regime in which the nonadiabatic excitations are strongly suppressed, so we disregard their effect on $S_z(t)$.  

For each spin, the probability of the nonadiabatic excitation can be 
estimated now from a spin-1/2 Hamiltonian (\ref{i-nlz}), in which  $S_z(t)$ is defined implicitly by Eq.~(\ref{sz-self}). 
Let us now change the time variable in (\ref{sz-self}) to 
$$
x=\beta t -kS_z(t),
$$
so that 
$$
x=\beta t + \frac{kx}{\sqrt{x^2+g^2}}.
$$
By differentiating, we find 
\begin{equation}
dt=\frac{dx}{\beta} \left(1- \frac{kg^2}{(x^2+g^2)^{3/2}} \right).
\label{diff1}
\end{equation}
Using this, we change the variables, from $t$ to $x$, in the time-dependent Schr\"odinger equation
$$
i\frac{d}{dt} |\psi \rangle = H_i(t) |\psi \rangle,
$$
where $H_i$ is defined in (\ref{i-nlz}) and write this equation in the matrix form 
\begin{equation}
  i\frac{d}{dx} \left(\begin{array}{c}
  a\\
  b
  \end{array}\right) = \alpha(x) \left( 
\begin{array}{cc}
x & g \\
g & -x
\end{array}
  \right) \left(\begin{array}{c}
  a\\
  b
  \end{array}\right),
    \label{se1}
\end{equation}
where $a$ and $b$ are the amplitudes of, respectively, the spin up and down states, and
\begin{equation}
\alpha(x)\equiv \frac{1}{\beta}\left(1- \frac{kg^2}{(x^2+g^2)^{3/2}} \right).
\label{al-def}
\end{equation}

Equation~(\ref{se1}) is very similar to the Landau-Zener model but it has an additional $x$-dependent factor, $\alpha(x)$, in front of the 2$\times$2 Landau-Zener matrix Hamiltonian. This difference, however, is crucially important. 

\subsection{Critical point}
For $k>g$ there is a range of $x$ for which $\alpha(x)<0$. From Eq.~(\ref{diff1}) this means that the function $x(t)$ is no longer monotonically growing, and hence the change of the time-variable is ill-defined. This problem can be traced to the fact that for $k>g$ Eq.~(\ref{sz-self}) that defines $S_z(t)$ has more than one solution for some $t$. 

This means that the original model (\ref{nLZ-def}) for $k>g$ describes the passage through a 1st order phase transition at some  time moment. The nonadiabatic excitations in the nLZ model in that case are abundant and have been already well studied for the nLZ model \cite{niu-nlz}. 
Hence, here we consider only the situation with $k\le g$, which avoids this phase transition, and for which our mean-field approximations leading to Eq.~(\ref{se1}) in the quasi-adiabatic limit are justified. Our primary interest, however, will be in the  passage near the phase transition, which corresponds to $g-k\ll g$.
A special value, $k=k_c=g$, corresponds to the control protocol that merely touches the critical point during the time evolution. In this case, Eq.~(\ref{diff1}) still has a unique solution for $t\in (-\infty,+\infty)$ but $S_z(t)$ changes so quickly near $t=0$ that the adiabaticity conditions cannot be satisfied even for $\beta \rightarrow 0$. 

\subsection{Energy degeneracy}
The Hamiltonian in Eq.~(\ref{se1}) has eigenvalues
\begin{equation}
\varepsilon_{\pm}=\pm \alpha(x)\sqrt{x^2+g^2}.
\label{en-def}
\end{equation}
In the Dykhne's approach, the adiabatic evolution is extended to the complex time so that we pass through the exact eigenenergy degeneracy point, at which $\varepsilon_+=\varepsilon_-$. If $\alpha(x)$ in (\ref{se1}) were a constant, as in the standard Landau-Zener model ($k=0$),  this point would be the branch cut at $x=ig$. However, for $k>0$ at this point $\alpha(x)$ diverges. Instead,  the energy degeneracy is found now when the prefactor $\alpha(x)$ is zero. Let us introduce a parameter 
\begin{equation}
\kappa \equiv k/g,
\label{kappa-def}
\end{equation}
that characterizes the relative strength of the spin interactions. Then,  $\alpha(x_0)=0$ is satisfied at
\begin{equation}
    x_0=ig\sqrt{1-\kappa^{2/3}}.
    \label{x0-def}
\end{equation}
For the moderate ferromagnetic interactions, $1>\kappa>0$, this zero is closer to the real axis than the point of the 
branch cut of the linear Landau-Zener model, at $x=ig$. This is the sign for the enhanced nonadadiabatic effects. At $\kappa=1$, which is the critical value, the degeneracy point touches the real time axis. 

Since near $x=x_0$ the adiabaticity is broken, we expect that the probability of the nonadiabatic excitation, which in our case is to remain in the spin up state as $t\rightarrow +\infty$, is given by  the adiabatic evolution along the time contour that passes through the degeneracy point:
\begin{equation}
P_{ex} \propto e^{-2{\rm Im} \left( 
\int_{0}^{x_0} dx \, \{\varepsilon_+(x)-\varepsilon_{-}(x) \}\right)}.
\label{dykhne1}
\end{equation}
Equation~(\ref{dykhne1}) is known as the Dykhne formula \cite{dykhne}.
Dykhne originally considered the typical situation with real parameters and  the energy difference in (\ref{dykhne1}) having a branch cut at $x=x_0$, as in the standard Landau-Zener Hamiltonian ($k=0$) for $x\approx x_0$: $\varepsilon_+-\varepsilon_{-} \propto \sqrt{x-x_0}$. For such cases, not only the leading exponent was found. Dykhne proved that the exponential prefactor in Eq.~(\ref{dykhne1}) at the leading order in small $\beta$  is generally $1$, as confirmed e.g. by the solution of the Landau-Zener model.    

In our case,  there is a complication. Although  $x=x_0$ is the point of the energy degeneracy, this degeneracy is not a branch cut point. Rather the entire Hamiltonian
\begin{equation}
  H(x) =\alpha(x) \left( 
\begin{array}{cc}
x & g \\
g & -x
\end{array}
  \right)
    \label{ham1}
\end{equation}
is a simple zero at this point. Namely, for $x\rightarrow x_0$,
$$
H(x) \propto (x-x_0). 
$$
The energy difference for $H(x)$ at $x_0$ has also a similar simple zero.  By investigating Eq.~(\ref{en-def}), with $\alpha(x)$  given in Eq.~(\ref{al-def}),  near $x=x_0$ given in Eq.~(\ref{x0-def}), we find
\begin{widetext}
\begin{equation}
 \varepsilon_+-\varepsilon_{-} =\frac{1}{\beta} \left\{ \frac{6i\sqrt{1-\kappa^{2/3}}}{\kappa^{1/3}}(x-x_0) +\frac{3(3-2\kappa^{2/3})}{\kappa g}(x-x_0)^2+\ldots \right\}.
    \label{endif}
\end{equation}
\end{widetext}

The effect of more complex degeneracy points on the Dykhne formula
was explored by Joye in \cite{joye-prj}. His theory says that if near the energy degeneracy point the energy difference behaves as $\varepsilon_+-\varepsilon_{-} \propto (x-x_0)^{n/2}$, and the entire Hamiltonian behaves as $H(x) \propto (x-x_0)^{m}$, then the exponent in the Dykhne formula has a prefactor
\begin{equation}
 4\sin^2\left(\frac{\pi(n-2m)}{2(n+2)} \right).
    \label{joye}
\end{equation}
Given Eq.~(\ref{endif}), we have $n=2$ and $m=1$, so the Joye's formula predicts the  zero prefactor, which would mean no nonadiabatic transitions. However, Joye  interpreted this fact in \cite{joye-prj}, so that a further theory was needed for this particular case. Hence,  we are to develop this theory in order to derive the probability of the nonadiabatic excitations for the evolution with the Hamiltonian (\ref{ham1}). 

\subsection{Adiabatic basis}
We will need the Schr\"odinger equation with $H(x)$ in the adiabatic basis. Let $|+\rangle$ and $|-\rangle$ be the eigenstates that correspond to the $\varepsilon_{\pm}$ eigenvalues. This basis is $x$-dependent, so after switching the basis, the adiabatic eigenstates of $H(x)$ are coupled by the time derivative: $i\langle +|\partial_x |-\rangle =-ig/[2(x^2+g^2)]$, so the Schr\"odinger equation is given by
\begin{equation}
i\frac{d}{dx}|\psi\rangle =
\left( 
\begin{array}{cc}
\varepsilon(x)/\beta & \frac{-ig}{2(x^2+g^2)} \\
\frac{ig}{2(x^2+g^2)} & -\varepsilon(x)/\beta
\end{array}
  \right) |\psi\rangle,
    \label{adiab}
\end{equation}
where 
$$
\varepsilon(x) \equiv \sqrt{x^2+g^2} - \frac{\kappa g^3}{x^2+g^2}.
$$
We assume that we start as $x\rightarrow -\infty$ in the ground state, so in this basis
$$
|\psi(-\infty) \rangle \sim \left(\begin{array}{c}
0\\
1
\end{array}\right)e^{-i\int_{-\infty}^{x}\varepsilon_{-} \,dx},
$$
and as $x\rightarrow +\infty$ the state behaves as 

\begin{eqnarray}
\nonumber |\psi(+\infty) \rangle \sim \left(\begin{array}{c}
B_+\\
0
\end{array}\right)e^{-i\int_{-\infty}^{x}\varepsilon_+(\tau) \,d\tau}+ \\
\left(\begin{array}{c}
0\\
B_{-}
\end{array}\right)e^{-i\int_{-\infty}^{x}\varepsilon_{-} (\tau) \,d \tau},
\end{eqnarray}
where $B_{\pm}$ are still unknown amplitudes and $\varepsilon_{\pm}=\pm \varepsilon/\beta$. The excitation probability is identified with $|B_+|^2$.

Note that the coefficient $1/\beta$ is large because we are interested in the quasi-adiabatic transition, which corresponds to a small transition rate $\beta$. This
makes the diagonal part of the Hamiltonian in (\ref{adiab}) generally large. The region with the nonadiabatic transitions then is crossed during a short time, which becomes shorter with decreasing $\beta$. The transition probability between the adiabatic states can then be, naively, estimated using the standard lowest nonzero order Born approximation. According to it, we treat the off-diagonal part of the matrix (\ref{adiab}) as a small perturbation and calculate the amplitude of the initially empty state up to the first order in this perturbation. The excitation probability is the square of this amplitude:
\begin{equation}
P_{ex} \approx \left| \int_{-\infty}^{\infty} dx\, \frac{g}{2(x^2+g^2)} e^{\frac{2i}{\beta}\int_{-\infty}^x  \varepsilon(\tau) \, d\tau} \right|^2.
\label{Born1}
\end{equation}
It is well known that such a Born approximation to $P_{ex}$ generally fails to predict the excitation probability when the latter is suppressed exponentially, e.g., as $P_{ex}\sim e^{-f/\beta}$. The reason is that the standard Born series is intended to produce  the contributions to $P_{ex}$ as powers of $\beta$. Therefore, formally different order terms in the Born series have to be suppressed additionally and thus can produce comparable contributions to  $P_{ex}$. 

The break down of the lowest Born approximation can be illustrated within the standard Landau-Zener model ($k=0$) for which the exponential prefactor in (\ref{Born1}) at the saddle point of the integral, at $x=ig$, diverges. By calculating such an integral, one would find a prefactor to the exponent in (\ref{plz})  different from the exactly known value, $1$. 
However,  in the context of the properties of the point $x_0$ that we encounter here, there is no such a singularity, so this formula is justified when it is applied to a piece of the integration time-path that goes through $x_0$.

\subsection{Touching the critical point}
Consider first the situation with touching the critical point, which takes place for $\kappa =1$ at $x_0=0$. The nonadiabatic transitions are mainly produced near the energy degeneracy point, so we can safely approximate near this point
$
\varepsilon (x)\approx \frac{3x^2}{2g} +O(x^3), 
$
and $g/(x^2+g^2) \approx 1/g+O(x^2)$. The Schr\"odinger equation then simplifies to
\begin{equation}
i\frac{d}{dx}|\psi\rangle =
\left( 
\begin{array}{cc}
\frac{3x^2}{2\beta g} & \frac{-i}{2g} \\
\frac{i}{2g} & -\frac{3x^2}{2\beta g}
\end{array}
  \right) |\psi\rangle.
    \label{adiab-2}
\end{equation}
The lowest Born approximation then predicts
\begin{equation}
P_{ex}^{(\kappa=1)} = \frac{1}{4g^2}\left|\int_{-\infty}^{\infty} dx\, e^{ix^3/(\beta g)}  \right|^2=\frac{\Gamma^2\left(\frac{1}{3}\right)}{12}\left(\frac{\beta}{g^2}\right)^{2/3},
    \label{pex2}
\end{equation}
where $\Gamma(\ldots)$ is the Euler's Gamma-function.
The lowest order approximation leads in (\ref{pex2}) to the power law, $P_{ex}\sim \beta^{2/3}$, rather than the exponential suppression of $P_{ex}$. Therefore, this approximation was well justified, and the higher  order terms in the Born series would produce only the corrections of higher power in $\beta$. Therefore, the numerical coefficient  
$$
\frac{\Gamma^2\left(\frac{1}{3}\right)}{12} \approx  0.698059
$$
can be trusted. 

\begin{widetext}

\begin{figure}[t!]
\centering \includegraphics[width=6in]{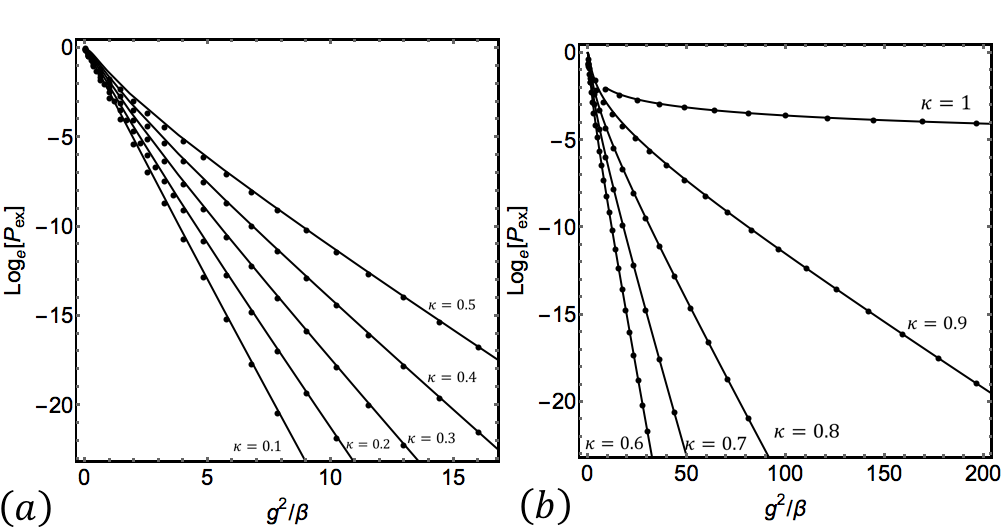}
\caption{Dependence of the excitation probability per spin in nLZ model on the dimensionless Landau-Zener parameter $g^2/\beta$. Different solid curves are our theoretical predictions for different  $\kappa\equiv k/g$, namely, Eq.~(\ref{pex2}) is for the curve at $\kappa=1$ and Eq.~(\ref{Pex3-2}) is for all $0<\kappa<1$. Here,
(a) $\kappa = 0.1,\, 0.2,\, 0.3,\, 0.4, \, 0.5$, and (b) $\kappa=0.6,\, 0.7,\, 0.8,\, 0.9,\, 1$. In all cases, the increase of $\kappa$ leads to a monotonic increase of $P_{ex}$.
The dots are the results of the numerical solution of the Schr\"odinger equation (\ref{adiab})  for time-evolution $t \in (-700,700)$. The prefactor is manifested as the deviation of the curves from the straight lines originating from $(0,0)$ point. The agreement of the numerically exact solution with theory is established when the prefactor becomes much smaller than $1$. Otherwise, small deviations from the analytical curves are visible. 
 }\label{Pex-fig}
\end{figure}
\end{widetext}
In Fig.~\ref{Pex-fig}, we compare  the numerically obtained solution for $P_{ex}$ (black dots) with our analytical predictions. In particular, in Fig.~\ref{Pex-fig}(b), the solid curve for $\kappa=1$ represents the prediction of the formula (\ref{pex2}).  The  agreement with the numerical  results is excellent.

Finally, we note that in the same regime Ref.~\cite{niu-nlz} predicted $P_{ex}\sim \beta^{3/4}$. The value $3/4$ is close to our $2/3$ but different. This discrepancy may be the result of the difference in the way we developed the mean field approximation. 

\subsection{Exponential suppression of the excitation probability}

For the differential equation (\ref{se1}), the  piece of the imaginary time axis that goes from zero to the branch cut at $x=ig$ is the Stokes line, along which the solution amplitudes are changing exponentially quickly. The strict adiabatic approximation generally fails near this line. One possibility to derive the transition amplitude between the adiabatic states then is to identify the region at the Stokes line  such that the behavior of the time-evolution across it simplifies. Inside this region, the evolution can be described analytically beyond the adiabatic approximation, and then it can be smoothly connected to the adiabatic states from both sides of the Stokes line \cite{joye-prj}. In our case, this region is found near the energy degeneracy point at $x=x_0$.    

\begin{figure}[t!]
\centering \includegraphics[width=\columnwidth]{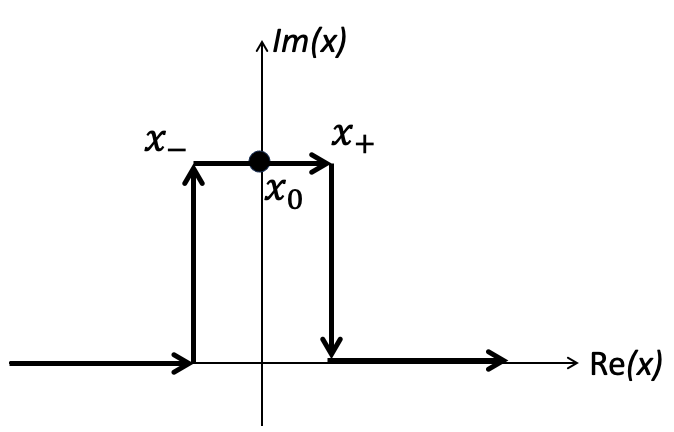}
\caption{The integration path (thick arrows) that starts and ends at real time infinities, and passes through the  energy degeneracy imaginary point $x_0$ along the line parallel to the real time axis. }\label{path-fig}
\end{figure}

 Equation~(\ref{se1}) has a single analytical solution on the complex plane with branch cuts starting at $x=\pm ig$ and going to the infinity along the imaginary time axis. Therefore, it is safe to deform the integration path from the ${\rm Re}(x)$-axis to the path shown by thick arrows in Fig.~\ref{path-fig}. The points $x_{\pm}$ along this path have the same imaginary value as the point $x_0$ in (\ref{x0-def}). The integration path turns from the real axis to reach the point $x_{-}$ along the vertical arrow. Then it goes through the point $x_0$ towards $x_{+}$, and then returns to the real time axis.

The vertical pieces are chosen to be sufficiently far from the point $x_0$, so that one can apply the adiabatic approximation. For the points  that lie on the path before  $x_{-}$ this leads us to an adiabatic amplitude
$|\psi(x)\rangle \sim e^{-i \int^{x}_{-\infty}  \varepsilon_{-}(\tau)\,d\tau} $. After passing through the degeneracy point, another asymptotic solution that behaves as $\sim e^{-i \int^{x}_{x_+}  \varepsilon_+(\tau)\,d\tau} $ acquires a finite amplitude, and its magnitude at the real axis is what determines the final excitation probability.

Along the arrow that connects $x_{-}$ and $x_+$, the adiabaticity is broken, so  our goal now is to find the absolute value squared of the amplitude of the new asymptotic solution that emerges by reaching the point $x_+$. 
By linearizing the time-dependence of the Hamiltonian parameters in Eq.~(\ref{adiab}) near $x_0$ (see the first term in Eq.~(\ref{endif})) in the adiabatic basis and setting $\tau\equiv x-x_0$ we find the effective Hamiltonian along the arrow that points to $x_+$: 
$$
H(\tau)\! = \!\left(
\begin{array}{cc}
\frac{3i\tau\sqrt{(1/\kappa)^{2/3}-1} }{\beta} &\frac{ -i}{2g\kappa^{2/3}}  \\
\frac{i}{2g\kappa^{2/3}}  & -\frac{3i\tau\sqrt{(1/\kappa)^{2/3}-1} }{\beta}
\end{array}
\right).
$$
Due to large values of $\beta$, the adiabaticity is violated in 
a very short region near $x_0$, so it is safe to assume that along the horizontal arrow in Fig.~\ref{path-fig} the time $\tau$ changes in the interval $\tau\in (-\infty,+\infty)$. 
Although $H(\tau)$ is non-Hermitian, the perturbative analysis applies to it equally well, as for the Hermitian evolution.  This Hamiltonian has the same property as the one from Eq.~(\ref{adiab-2}), for the case with $\kappa=1$. Namely, along the integration path, at $\tau=0$, the energy gap almost closes, while the off-diagonal terms are non-singular at $x_0$. Hence, the transition amplitude
from the initial adiabatic state at the point $x_{-}$
to the new state  with the growing  phase asymptotic $\sim e^{-i\int^x \varepsilon_+(\tau)\, d\tau}$ 
can be estimated safely, up to an unimportant unitary phase factor, by applying the lowest Born approximation: 
$$
A=\frac{1}{2g\kappa^{2/3}} \int_{-\infty}^{\infty} d\tau \,
e^{-\frac{3\tau^{2}}{\beta} \sqrt{(1/\kappa)^{2/3}-1} }.
$$
Finally, along the down-pointing arrow in Fig.~\ref{path-fig}, this new state gains an additional adiabatic phase factor $ e^{-i\int_{x_{+}}^x d\tau\, \varepsilon_+(\tau)}$.

Due to the large values of $1/\beta$, the adiabaticity is broken only in the vanishing with $\beta$  vicinity of $x_0$. Hence, $x_{\pm}$ in this limit can be placed arbitrarily close to $x_0$. Then, for the adiabatic evolution along the vertical arrows, it is safe to disregard the real components of $x_+$ and $x_{-}$, so that the contribution of the vertical parts of the integration contour produce the standard Dykhne exponent (\ref{dykhne1}) factor in the excitation probability.   The horizontal path that goes through $x_0$ contributes with an additional $|A|^2$ factor to this probability. By collecting the contributions from all pieces of the integration contour we find

\begin{equation}
P_{ex}=|A|^2e^{-2{\rm Im} \left(\int_{0}^{x_0} dx\, (\varepsilon_+(x)-\varepsilon_{-}(x)) \right)}.
\label{Pex3}
\end{equation}
All integrals in (\ref{Pex3})
 can be calculated analytically, so we finally find

\begin{equation}
P_{ex}=\frac{\pi}{12\kappa\sqrt{1-\kappa^{2/3}}} \left(\frac{\beta}{g^2} \right) 
\exp\left(-\frac{2g^2}{\beta} F(\kappa)
\right),
\label{Pex3-2}
\end{equation}
where
 \begin{widetext}     
\begin{equation}
F(\kappa)\equiv \kappa^{1/3}\sqrt{1-\kappa^{2/3}} +{\rm arccos}(\kappa^{1/3}) +\kappa \log_e\left( \frac{2-2\sqrt{1-\kappa^{2/3}}-\kappa^{2/3}}{\kappa^{2/3}} \right),
\label{fk}
\end{equation}
 \end{widetext}
 and where $\kappa$ is defined in terms of the parameters of the nLZ model (\ref{nLZ-def}) by Eq.~(\ref{kappa-def}). 
 Equations~(\ref{Pex3-2}) and (\ref{fk}) are our main results. They show explicitly that the excitation probability has the functional form $P_{ex}\propto \beta e^{-f/\beta}$, i.e., it has a nontrivial prefactor that decays to zero as $\beta^{\nu}$. Interestingly, the exponent $\nu =1$ in the   prefactor in (\ref{Pex3-2}) is different from the exponent $\nu_c=2/3$ for the case of touching the critical point at $\kappa=1$. Mathematically, this follows from Eq.~(\ref{endif}). Namely, for $1>\kappa>0$ the leading term in the energy difference near $x=x_0$ is growing linearly with $x-x_0$ but for $\kappa=1$ this term is identically zero, so the next order in $(x-x_0)$ term controls the energy splitting.

\begin{figure}[t!]
\centering \includegraphics[width=3in]{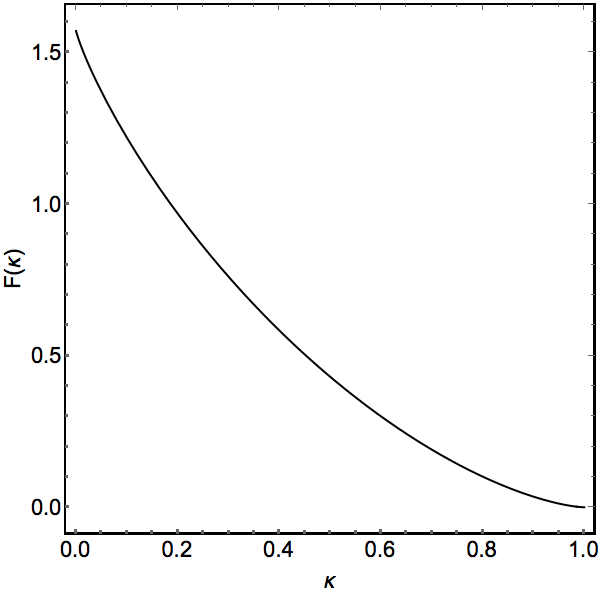}
\caption{The factor from the exponent in Eq~(\ref{Pex3-2}), $F(\kappa)$, given by Eq.~(\ref{fk}). }\label{fk-fig}
\end{figure}
The function $F(\kappa)$ is shown in Fig.~\ref{fk-fig}. As $\kappa\rightarrow 0$, it approaches the limit
$$
F(0)=\pi/2,
$$
which reproduces the exponent in the Landau-Zener formula (\ref{plz}). Near the critical point, $1-\kappa \ll 1$, it behaves as a power law 
\begin{equation}
F(\kappa) \approx \frac{4\sqrt{2}}{3\sqrt{3}} (1-\kappa)^{3/2}.
\label{exp-exp}
\end{equation}
As expected, $F(1)=0$, which means a transition to the regime with only a power law suppression of the nonadiabatic excitations. Our expression for $F(k)$ differs quantitatively from the one that was derived in Ref.~\cite{niu-nlz} for the nLZ model. Thus, instead of the exponent $\mu=3/2$ in (\ref{exp-exp}), Ref.~\cite{niu-nlz} predicted $\mu=2$.  Nevertheless, qualitatively our Fig.~\ref{fk-fig} is very similar to its analog, Fig.~5 in \cite{niu-nlz}. Within our mean field approximation, however, we achieved an excellent agreement with exact numerical simulations.

 For comparison with numerical results, we noted that Eq.~(\ref{Pex3-2}) was derived for the limit $\beta/g^2\ll 1$, so it made unphysical predictions for the strongly nonadiabatic regime $\beta/g^2\gg 1$. To avoid this complication, we replaced the prefactor
 $$
 \frac{\pi}{12\kappa\sqrt{1-\kappa^{2/3}}} \left(\frac{\beta}{g^2} \right) 
 $$
 by the function
$$
1-\exp\left(-\frac{\pi}{12\kappa\sqrt{1-\kappa^{2/3}}} \left(\frac{\beta}{g^2} \right) \right). 
$$
This did not spoil the asymptotic behavior of the prefactor as $\beta\rightarrow 0$, but interpolated it to a known value $P_{ex}(\beta \rightarrow \infty)=1$ in order to avoid unphysical predictions with $P_{ex}>1$.

In Fig.~\ref{Pex-fig}, we compare such adjusted predictions of Eq.~(\ref{Pex3-2}) to the result of the exact numerical integration of the evolution equation~(\ref{adiab}), which is equivalent to the original Schr\"odinger equation (\ref{se1}). The agreement is excellent even for the small values of $\kappa$, which are far from the critical point $\kappa_c=1$. It seems that for $0<\kappa<1$ the only condition for the validity of  Eq.~(\ref{Pex3-2}) is the smallness of the prefactor: $|A|^2\ll 1$, which is guaranteed at sufficiently small ratio, $\beta/g^2$. However, for $\kappa \ll 1$, this is achieved for relatively small $P_{ex}$. Hence, for the passage at a larger distance from the critical point the role of the prefactor is smaller.

Finally, we note that our analysis is straightforward to extend to the antiferromagnetic interactions with $\kappa<0$ but the analytical formulas become much more complex. For $\kappa<0$ there are two rather than one zeros of the function $\alpha (x)$ at the same distance from the real time axis. The vicinity of each zero can be treated similarly to the zero at $x_0$ for $\kappa>0$. However, in addition, we then must take into account the interference between the contribution of each zero to the transition amplitude. This leads to oscillations in the exponential prefactor. As this regime is far from the critical point $\kappa_c=1$, we leave it without further discussion. 

\section{Solvable model: driven Stark ladder}

Although the nLZ model is  specific, we conjecture that our qualitative observations about the behavior of the exponential prefactor are universal for quantum critical phenomena. Thus, for many interacting spin systems with a field ramp near the critical point, the mean field approximation should generally lead to a two-state evolution with a time-dependent prefactor that renormalizes the relevant interactions in the vicinity of the critical point. Similarly to the nLZ model, this factor should generally have  a simple zero that should determine the position of the energy level crossing in the complex time-plane. As we have shown for the nLZ model, in such situations the excitation probability is expected to acquire a nontrivial power-law prefactor at the Dykhne's exponent.

It is interesting whether our conjecture can be extended beyond the mean field theory. To support it, we now work out a simple model that describes a passage near a critical point without approximations, such as the mean field approximation that we had to use for the nLZ model. 

The following model has a critical point not in the sense of the discontinuity of the ground state of a many-body Hamiltonian. Rather it corresponds to an isolated critical point that separates phases with quantum localization. Nevertheless, this model has a well-defined thermodynamic limit, $N\rightarrow \infty$, where $N$ is the number of the interacting states. Hence, the phenomenology of the dynamic quantum phase transitions applies to it. The advantage of this model is that it is completely solvable analytically for any driving protocol. 

\subsection{Stark ladder Hamiltonian}

\begin{widetext}

\begin{figure}[t!]
\centering \includegraphics[width=5in]{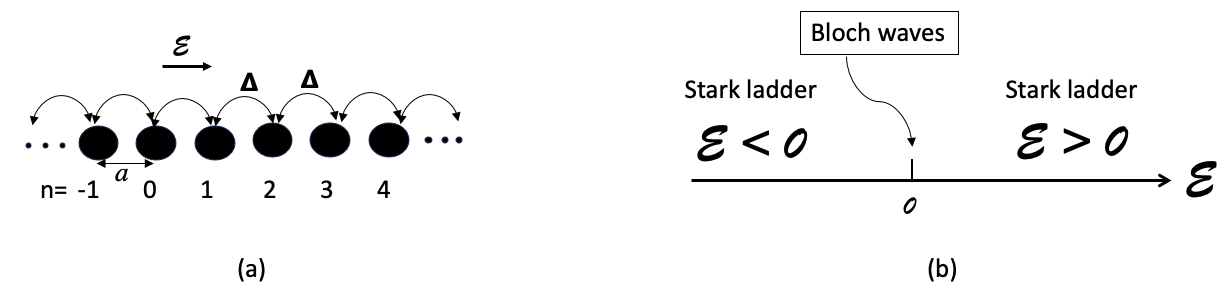}
\caption{(a) The tight-binding model of a quantum particle hopping with an amplitude $\Delta$ between neighboring nodes of a linear chain. The electric field ${\cal E}$ is applied along the chain to split the potential energy of the particle on the nodes, as the nearest nodes  are separated by the distance $a$. (b) The phase diagram showing that for both ${\cal E}>0$ and ${\cal E}<0$ the eigenstates of the Hamiltonian are the spatially localized Stark states, whereas ${\cal E}=0$ is the critical point at which the eigenstates are the delocalized Bloch waves.}
\label{stark-fig}
\end{figure}
\end{widetext}

Consider a model of a linear chain of sites with quantum coherent hopping of a charged particle.  The chain is placed in an external constant electric field. 
Let $|n\rangle$, where $n=-\infty,\ldots, \infty $, be orthogonal states of the particle, with  spatially separated average  positions. Let   $a$ be the distance between the sites $n$ and $n\pm 1$, as shown in Fig.~\ref{stark-fig}(a). The direct hopping between  the neighboring states has the energy amplitude $\Delta$. 

The electric field  creates a linear potential for the particle with electric charge $e$ along the  $x$-axis of the chain: $U(x)=e{\cal E}x$. In the tight-binding model, this corresponds to the energy $e{\cal E}an$ for the particle in the $n$-th node of the chain.  Thus, the entire tight-binding Hamiltonian is given by
\begin{equation}
    H=b n|n\rangle\langle n| +\Delta (|n\rangle \langle n+1| +|n+1\rangle\langle n|), \quad b \equiv e{\cal E}a.
    \label{simple-chain}
\end{equation}

For any finite electric field, and in the absence of decoherence/relaxation, the eigenstates of the Hamiltonian (\ref{simple-chain}) are the localized Stark states \cite{sinitsyn-prb2002}, which have a spectrum with a fixed energy gap between the nearest energy levels. Such a spectrum is called the Stark ladder. Nevertheless, at ${\cal E}=0$, the eigenstates are the delocalized Bloch waves (see Fig.~\ref{stark-fig}(b)). Hence,  $b=0$ in (\ref{simple-chain}) is the critical point. If the electric field changes sufficiently slowly without crossing the value ${\cal E}=0$, the predictions of the adiabatic theorem are satisfied and we expect to find an exponential suppression of the excitations. However, if we cross the point ${\cal E}=0$ we expect an abundant production of the excitations even for a quasi-adiabatic evolution. 

\subsection{The number of nonadiabatic excitations}

Let the time-dependent electric field  start and end at large absolute values: $|b|_{t\rightarrow \pm \infty}\gg \Delta$ in (\ref{simple-chain}). Then, the Stark states are asymptotically the basis states $|n\rangle$. Hence, the number of the excitations can be identified with the change of the index $n$. Let $n_0$ be the initial node of the particle as $t\rightarrow -\infty$ and let as $t\rightarrow +\infty$ there is a probability distribution $P(n)$ to find the particle at the $n$-th node. 
On the Stark ladder, between the energies of the Stark states $|n\rangle$ and $|n_0\rangle$ there are $(n-n_0)$ elementary energy gaps.   Hence, it is natural to define the average number of the nonadiabatic excitations as
$$
n_{ex} \equiv \sqrt{\langle (n-n_0)^2 \rangle},
$$
where the average is over the final distribution $P(n)$.

The linear dynamic transition through the critical point in the model (\ref{simple-chain}), namely the case with $b=\beta t$, was previously studied in \cite{sinitsyn-prb2002}. As  for other critical phenomena, the exact solution of this model produced a power law for the number of the excitations:
 \begin{equation}
n_{ex} =\sqrt{\frac{4\pi  \Delta^2}{\beta}} \propto \beta^{-1/2}. 
    \label{nav-houston}
\end{equation}

Let us now consider a different protocol.
In what follows, we are interested in the Hamiltonian (\ref{simple-chain}) with 
\begin{equation}
b(t)=(\varepsilon t)^2+g, \quad \varepsilon,g>0.
\label{protocol}
\end{equation}
Here, the electric field becomes infinitely large as $t\rightarrow \pm \infty$. However, it is  never zero and  reaches the minimum $b=g$ at $t=0$. The rate of the passage through this minimum is controlled by the parameter $\varepsilon$. Our goal is to find the scaling of $n_{ex}(\varepsilon)$ for $\varepsilon \rightarrow 0$ after the time evolution during $t\in (-\infty,+\infty)$.

\subsection{The amplitude generating function} 
 We will search for the solution to the Schr\"odinger equation of the form
\begin{equation}
|\psi(t)\rangle = \sum_{n=-\infty}^{\infty} a_n (t) |n\rangle, 
    \label{anz1}
\end{equation}
and introduce the {\it amplitude generating function} (AGF)
\begin{equation}
    u(t;\phi)\equiv \sum_{n=-\infty}^{\infty} a_n (t)e^{in\phi},
    \label{agf1}
\end{equation}
from which we can find all state amplitudes in the original basis by the inverse Fourier transformation:
\begin{equation}
    a_n(t)= \frac{1}{2\pi}\int_0^{2\pi}  e^{-in\phi} u(t;\phi) \, d\phi.
    \label{amp1}
\end{equation}
The AGF is also convenient for finding the moments of the position index $n$. Thus, if we normalize 
\begin{equation}
    \frac{1}{2\pi}\int_0^{2\pi} d\phi \,|u(t;\phi)|^2 = 1,
    \label{agf-norm}
\end{equation}
then the average of the $m$-th power of  $n$ is given by
\begin{equation}
  \langle n^m(t) \rangle = 
  \frac{1}{2\pi} \int_0^{2\pi}
  d\phi \, u^*(t;\phi) \left(-i \frac{d}{d\phi} \right)^m u(t;\phi), \quad m=1,2,\ldots.
    \label{avm}
\end{equation}

 \subsection{Exponentially suppressed nonadiabatic excitations} 
 The time-dependent Schr\"odinger equation for the state amplitudes is
\begin{equation}
    i\dot{a}_n=(\varepsilon^2 t^2+g)n a_n +\Delta(a_{n-1}+a_{n+1}), \quad n=-\infty, \ldots, +\infty.
    \label{chain-t2}
\end{equation}
Changing the variables $a_{n} \rightarrow a_ne^{-i n\left( \frac{\varepsilon^2 t^3}{3} + gt\right)}$, then multiplying the $n$-th equation in (\ref{chain-t2}) by $e^{i\phi n}$, and summing all these equations, we find  the first order differential equation for the AGF, $u(t)\equiv u(t;\phi)$:

\begin{equation}
  i \frac{du(t)}{dt}=2\Delta \cos \left(\frac{\varepsilon^2 t^3}{3} +gt +\phi  \right)u(t), 
    \label{agf-c1}
\end{equation}
where $\phi\in[0,2\pi)$ should be treated as a constant parameter.

Let  as $t\rightarrow -\infty$ the system start with the particle at the node with $n=0$. The initial condition for Eq.~(\ref{agf-c1}) is then $u(\--\infty)=1$. The solution for the AGF as $t\rightarrow +\infty$ then is given by
\begin{equation}
    u(t=\infty) = \exp\left\{-2i\Delta \cos \phi \int_{-\infty}^{\infty} d\tau \cos \left( \frac{\varepsilon^2 \tau^3}{3} + g\tau \right) \right\}.
    \label{agf-airy}
\end{equation}

\begin{figure}[t!]
\centering \includegraphics[width=3in]{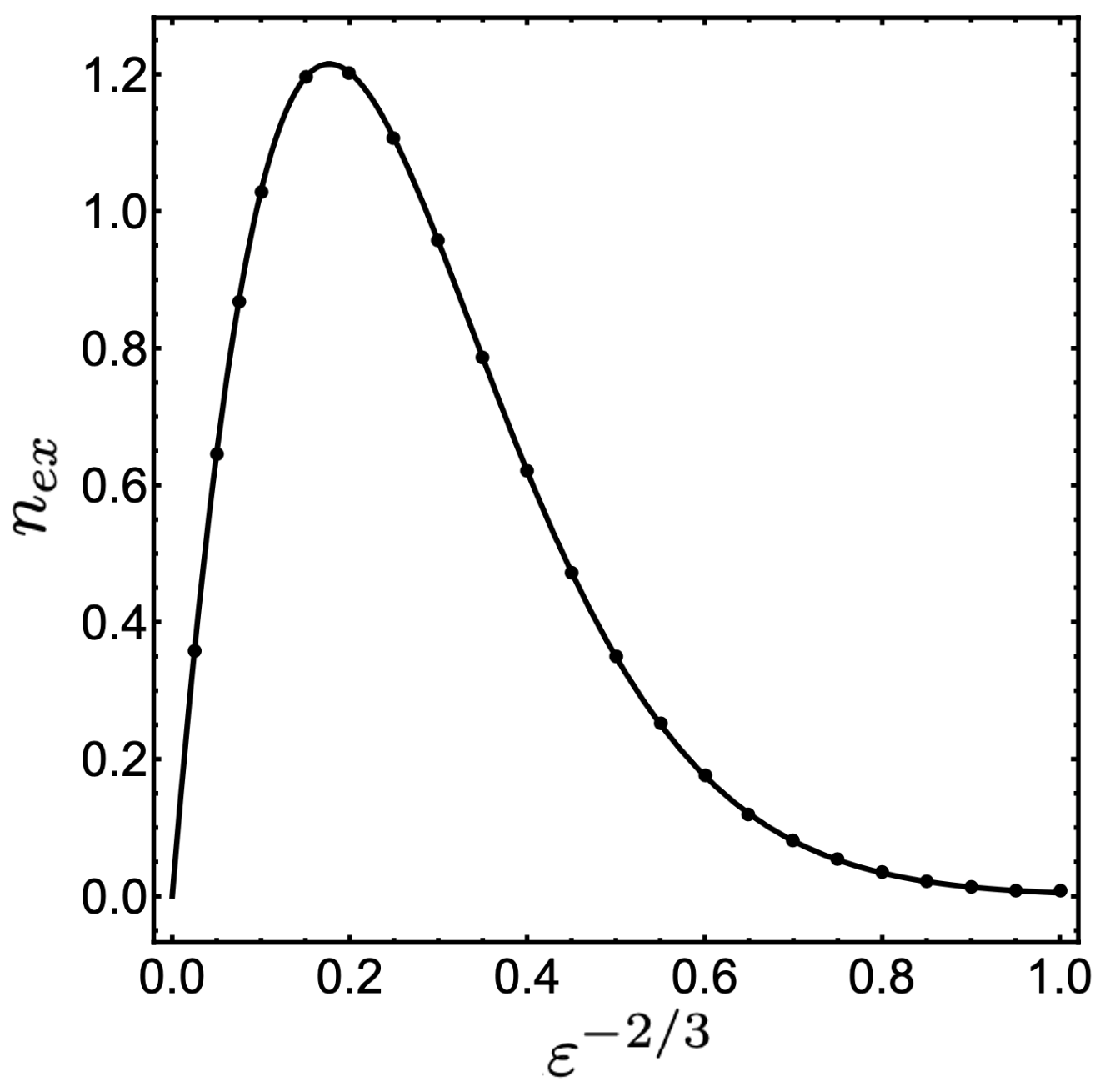}
\caption{The number of the excitations $n_{ex}$ for different characteristic transition times, $\varepsilon^{-2/3}$. The solid curve is  the theoretical prediction, Eq~(\ref{nex-quad1}). The dots are the results of the direct numerical solution of the Schr\"{o}dinger equation, Eq~(\ref{chain-t2}), for the time-evolution during $t \in (-50,50)$, starting in the middle of a chain with $50$ sites. Here, $\Delta=g=5$.}\label{nex-fig}
\end{figure}

 The  integral in (\ref{agf-airy})  can be expressed via the Airy function:
$$
{\rm Ai}[x] \equiv \frac{1}{\pi} \int_0^{\infty} dt\, \cos \left(\frac{t^3}{3} +xt \right),
$$
in terms of which  
\begin{equation}
    u(t=\infty) = \exp\left(-4\pi i\Delta\varepsilon^{-2/3} {\rm Ai}\left[ g\varepsilon^{-2/3} \right]\cos \phi  \right). 
    \label{agf-airy2}
\end{equation}
Using Eq.~(\ref{avm}), we get
\begin{equation}
   n_{ex} \equiv \sqrt{\langle n^2\rangle} = \sqrt{8}\pi\Delta \varepsilon^{-2/3} {\rm Ai}\left[ g\varepsilon^{-2/3} \right].
    \label{nex-quad1}
\end{equation}
Thus, the number of the nonadiabatic excitations, $n_{ex}$, up to a constant prefactor, is described by the function $x{\rm Ai}[x]$, where $x= g\varepsilon^{-2/3}$. Note that $\varepsilon^{-2/3}$ has the dimension of time. Large values of this time correspond to the quasi-adiabatic regime.  Figure~\ref{nex-fig},  confirms the validity of Eq.~(\ref{nex-quad1}) by comparing the analytical results with the numerical solution of the Schr\"odinger equation.

In the quasi-adiabatic limit the argument of the Airy function in (\ref{nex-quad1}) is large, so we can replace ${\rm Ai}[x]$ by its known asymptotic value
$$
{\rm Ai}[x] \sim\frac{1}{2\sqrt{\pi}x^{1/4}} e^{-\frac{2}{3}x^{3/2}}, \quad x\rightarrow \infty.
$$

Finally, we find for the quasi-adiabatic evolution
\begin{equation}
   n_{ex}  \approx \frac{\sqrt{2\pi}\Delta }{g^{1/4}\varepsilon^{1/2}} e^{-\frac{2 g^{3/2}}{3\varepsilon} }.
    \label{nex-quad2}
\end{equation}
This result shows explicitly that for $g>0$, i.e., as far as there is no crossing of the critical point by the external field, the number of the excitations is suppressed in the adiabatic limit  exponentially but with a nontrivial prefactor of the exponent.   This prefactor  scales as the power law, $ \varepsilon^{\nu}$, where $\nu=-1/2$, on the transition rate $\varepsilon$. Equation~(\ref{nex-quad2}) also reveals the ``exponent-in-exponent" behavior. Namely, in our case the parameter $g$ controls the minimal distance to the critical point  that is encountered during the process.  The factor in the exponent in (\ref{nex-quad2}) scales as  $\propto g^{\mu}$, where $\mu =3/2$. 

We also comment on the case of merely touching the critical point. This corresponds in our protocol to $g=0$.  
Equation~(\ref{nex-quad2}) at $g=0$ diverges, which suggests a possible new power law behavior of $n_{ex}$. Indeed, by setting $g=0$ in Eq.~(\ref{agf-airy}) we find
\begin{equation}
    n_{ex} = \frac{\sqrt{2}\Gamma\left(1/3\right)\Delta}{3^{1/6}\varepsilon^{2/3}} \approx 3.1547 \cdot \frac{\Delta}{\varepsilon^{2/3}}.
\end{equation}
This is the power law, $n_{ex}\propto \varepsilon^{\nu_c}$, with $\nu_c=-2/3$, which is different from $\nu=-1/2$ in the prefactor of Eq.~(\ref{nex-quad2}).

\section{Conclusion}

We solved two very different models that describe a slow passage of an explicitly time-dependent quantum system near a critical point. In one case the model was the nonlinear Landau-Zener model, which we initially simplified using the self-consistent mean field approximation and then applied the complex analysis in the spirit of the derivation of the Dykhne formula. The other case was a fully solvable model of a dynamic transition between the localized states in the Stark ladder after the ``dive" of the electric field to the vicinity of the critical point at zero field. 

Despite strong differences in the physics of these models and the methods that we applied, several findings turned out to be common  for both the nLZ  and the driven Stark ladder models:

(i) After the system passes near a critical point, with a characteristic rate $\beta$, the number of the nonadiabatic excitations scales with small $\beta$ as $P_{ex}\propto \beta^{\nu}e^{-f/\beta}$, with some exponent $\nu$.

(ii) Let $\kappa$ be the control parameter, such that $\kappa =\kappa_c$ marks the regime of touching the critical point. Then, near the critical $\kappa$, there is an ``exponent-in-exponent", $\mu$, such that $f\sim |\kappa-\kappa_c|^{\mu}$.

(iii) For $\kappa=\kappa_c$, the number of the produced excitations scales as a power law $P_{ex}\sim \beta ^{\nu_c}$, with $\nu_c$ different from $\nu$ in (i).

This similarity between the predictions of  two very different models strongly points to possible universality of the properties (i)-(iii) among many other quantum critical phenomena.

\begin{acknowledgements}
This work was supported in part by the U.S. Department of Energy, Office of Science, Office of Advanced Scientific Computing Research, through the Quantum Internet to Accelerate Scientific Discovery Program, and in part by U.S. Department of Energy under the LDRD program at Los Alamos. V.G.S. acknowledges funding support for travel from St. John's College, University of Cambridge, through a Graduate Scholars research expenses scheme.
\end{acknowledgements}


\end{document}